\documentclass[cits]{PoS}

\title{Electromagnetic effects in $\eta\to3\pi$}
\ShortTitle{Electromagnetic effects in $\eta\to3\pi$}
\author{\speaker{Christoph Ditsche}\\
        Helmholtz-Institut f\"ur Strahlen- und Kernphysik (Theorie) and
        Bethe Center for Theoretical Physics, Universit\"at Bonn, D-53115 Bonn, Germany\\
        E-mail: \email{ditsche@hiskp.uni-bonn.de}}
\author{Bastian Kubis\\
        Helmholtz-Institut f\"ur Strahlen- und Kernphysik (Theorie) and
        Bethe Center for Theoretical Physics, Universit\"at Bonn, D-53115 Bonn, Germany\\
        E-mail: \email{kubis@hiskp.uni-bonn.de}}
\author{Ulf-G. Mei{\ss}ner\\
        Helmholtz-Institut f\"ur Strahlen- und Kernphysik (Theorie) and
        Bethe Center for Theoretical Physics, Universit\"at Bonn, D-53115 Bonn, Germany\\
        Institut f\"ur Kernphysik (Theorie), Institute for Advanced Simulations and
        J\"ulich Center for Hadron Physics, Forschungszentrum J\"ulich, D-52425  J\"ulich, Germany\\
        E-mail: \email{meissner@hiskp.uni-bonn.de}}

\abstract{We re-evaluate the electromagnetic corrections to $\etp$ decays at next-to-leading order in the chiral expansion, arguing that effects of order $e^2(m_d-m_u)$ disregarded so far are not negligible compared to other contributions of order $e^2$ times a light-quark mass. Despite the appearance of the Coulomb pole in $\etpc$ and cusps in $\etpn$, the overall corrections remain small.}

\FullConference{6th International Workshop on Chiral Dynamics\\
		July 6-10 2009\\
		Bern, Switzerland}

\usepackage{amsmath,amsfonts}

\providecommand{\tn}[1]{\textnormal{#1}}
\providecommand{\ord}{\mathcal{O}}
\providecommand{\amp}{\mathcal{A}}

\providecommand{\me}{M_\eta^2}
\providecommand{\mpi}{M_\pi^2}
\providecommand{\mpc}{M_{\pi^\pm}^2}
\providecommand{\mpn}{M_{\pi^0}^2}

\providecommand{\etp}{\eta\to3\pi}
\providecommand{\etpc}{\eta\to\pi^0\pi^+\pi^-}
\providecommand{\etpn}{\eta\to3\pi^0}

\providecommand{\ev}{\,\tn{eV}}

\DeclareMathOperator{\ndf}{ndf}

\begin{document}

%================================================================================
\section{Introduction}
%\label{sec:intro}
%================================================================================

The decay $\etp$ is particularly interesting because it is forbidden by isospin symmetry and thus can only happen via isospin breaking due to either strong interactions,
\begin{equation}
 \mathcal{H}_\tn{QCD}(x)=\frac{m_d-m_u}{2}(\bar{d}d-\bar{u}u)(x)\;,
\end{equation}
which are proportional to the light-quark mass difference $m_d-m_u$, or electromagnetic interactions,
\begin{equation}
 \mathcal{H}_\tn{QED}(x)=-\frac{e^2}{2}\int dy\; D^{\mu\nu}(x-y)T(j_\mu(x)j_\nu(y))\;,
\end{equation}
which are proportional to the electric charge squared.
A well-known low-energy theorem by Sutherland~\cite{sutherland} states that the electromagnetic effects in $\etp$ are small. This decay is therefore very sensitive to $m_d-m_u$ and hence it potentially yields a particularly clean access to the determination of quark mass ratios. The systematic machinery that can cope with both effects accurately and leads to many experimentally testable predictions like Dalitz plot parameters, the decay width and quark mass ratios is chiral perturbation theory~(ChPT)~\cite{weinbergchpt,glannphys,glchpt} with the inclusion of electromagnetism~\cite{urech}. In view of new and upcoming high statistics experiments~\cite{nKLOEnew,CELSIUSWASA,cKLOEnew,WASAatCOSY,unverzagt,prakhov} we reconsider the electromagnetic corrections to these observables.

While early calculations of the \textit{strong} amplitude at tree level yielded a decay width which is off from the experimental value by a factor of a few, Gasser and Leutwyler~(GL)~\cite{gleta} calculated the strong contributions at one-loop level and observed large unitarity corrections due to strong final-state interactions (i.e.\ $\pi\pi$ rescattering), but their result still differs from experiment by a factor of roughly 2. Thus strong corrections beyond one loop were studied using dispersive methods~\cite{kambor,anisovich}, unitarized ChPT~\cite{beisert,borasoy}, and finally with a complete two-loop calculation~\cite{bijnens}. All of these found considerable enhancement compared to the one-loop calculation.

According to Sutherland's theorem, the \textit{electromagnetic} contributions at tree level vanish. Baur, Kambor, and Wyler~(BKW)~\cite{bkw} studied corrections to Sutherland's theorem by evaluating the electromagnetic effects at one-loop level, but they found them to be very small.
The motivation for reconsidering electromagnetic corrections at chiral order $p^4$ in the present work hinges on the fact that BKW neglected terms proportional to $e^2\delta$, where $\delta=m_d-m_u$, by arguing that these are of second order in isospin breaking and therefore expected to be suppressed even further. However, by restricting oneself to terms of the form $e^2m_s$ and $e^2\hat{m}$, where $\hat{m}=(m_u+m_d)/2$, one excludes some of the most obvious electromagnetic effects like real- and virtual-photon contributions as well as effects due to the charged-to-neutral pion mass difference $\Delta\mpi=\mpc-\mpn=\ord(e^2)$, both of which scale as $e^2\delta$.
These mechanisms fundamentally affect the analytic structure of the amplitudes in question: in the charged decay channel $\etpc$ there is a Coulomb pole at the $\pi^+\pi^-$ threshold $s=4\mpc$, while in the neutral decay channel $\etpn$ the pion mass difference induces a cusp behavior at this threshold due to $\pi^+\pi^-\to\pi^0\pi^0$ rescattering, compare e.g.\ Ref.~\cite{MMS}. In contrast, the corrections identified by BKW are all polynomials (due to counterterms) or quasi-polynomials (due to kaon loop effects) inside the physical region. Furthermore, Sutherland's theorem guarantees that at the soft-pion point these corrections scale as $e^2\hat{m}$ only (and not $e^2m_s$); hence the relative suppression of the neglected terms is of the order of $\delta/\hat{m}\approx2/3$ and therefore not a priori small.

The details of the presented analysis can be found in Ref.~\cite{dkm}.

%================================================================================
\section{$\boldsymbol{\eta\to}3\boldsymbol{\pi}$ decay amplitudes at $\boldsymbol{\ord(e^2\delta)}$}
%\label{sec:amplitudes}
%================================================================================

At leading chiral order the $\etp$ decay amplitudes for the charged and the neutral channel,
\begin{align}
 \begin{split}
 \langle\pi^0\pi^+\pi^-|\eta\rangle &= i(2\pi)^4\delta^4(p_{\pi^0}+p_{\pi^+}+p_{\pi^-}-p_\eta)\amp_c(s,t,u)\;,
 \\
 \langle3\pi^0|\eta\rangle &= i(2\pi)^4\delta^4(p_{\pi^0_1}+p_{\pi^0_2}+p_{\pi^0_3}-p_\eta)\amp_n(s,t,u)\;,
 \end{split}
\end{align}
follow from the tree graphs where the fields in the LO Lagrangian are diagonalized by use of the $\eta\pi^0$ mixing angle $\epsilon=(\sqrt{3}/4)\times\delta/(m_s-\hat{m})+\ord(\delta^3)$.
For the \textit{charged} channel, the Mandelstam variables $s=(p_\eta-p_{\pi^0})^2$, $t=(p_\eta-p_{\pi^+})^2$ and $u=(p_\eta-p_{\pi^-})^2$ are related by $s+t+u=$\linebreak$\me+\mpn+2\mpi=3s_0^c$. Expanding in isospin breaking parameters up to $\ord(\delta,e^2,e^2\delta)$ and rewriting everything in terms of physical observables like $\me$, $\mpn$, $\Delta\mpi$ and the pion decay constant $F_\pi$ yields for the charged amplitude at leading chiral order
\begin{equation}\label{eqn:cloamp}
 \amp_c^\tn{LO}=-\frac{B_0\delta}{3\sqrt{3}F_\pi^2}\left\{1+\frac{3(s-s_0^c)+2\Delta\mpi}{\me-\mpn}\right\}=-\frac{(3s-4\mpn)(3\me+\mpn)}{Q^216\sqrt{3}F_\pi^2\mpn}\;.
\end{equation}
An additional electromagnetic term of order $e^2\delta$ cancels the pion mass difference that is implicitly included in $s_0^c$. The LO charged amplitude is completely proportional to $\delta$ and can thus be expressed in terms of the quark mass double ratio $Q^2=(m_s^2-\hat{m}^2)/(m_d^2-m_u^2)$ which is particularly stable with respect to strong higher-order corrections~\cite{glchpt,leuellipse}. The amplitude depends linearly on $s$, and by inserting $s_0^c$ it explicitly displays the Adler zero at $s=4\mpn/3$.
For the \textit{neutral} channel, the Mandelstam variables $s=(p_\eta-p_{\pi^0_1})^2$, $t=(p_\eta-p_{\pi^0_2})^2$ and \mbox{$u=(p_\eta-p_{\pi^0_3})^2$} are related by \mbox{$s+t+u=\me+3\mpn=3s_0^n$}. The neutral amplitude at leading chiral order,
\begin{equation}\label{eqn:nloamp}
 \amp_n^\tn{LO}=-\frac{B_0\delta}{\sqrt{3}F_\pi^2}=-\frac{3(\me-\mpn)(3\me+\mpn)}{Q^216\sqrt{3}F_\pi^2\mpn}\;,
\end{equation}
also carries an overall factor of $\delta$, but it contains neither derivatives nor electromagnetic terms and hence it is just a constant.

While for the charged decay invariance under charge conjugation implies that the amplitude $\amp_c(s,t,u)$ is symmetric under the exchange of $t$ and $u$, the amplitude $\amp_n(s,t,u)$ for the neutral decay has to be symmetric under exchange of all pions and thus all Mandelstam variables due to Bose symmetry. All calculations of both $\etp$ decay channels performed in ChPT so far used the following relation between the charged and the neutral amplitude that utilizes isospin symmetry,
\begin{equation}\label{eqn:isoamprel}
 \amp_n(s,t,u)=\amp_c(s,t,u)+\amp_c(t,u,s)+\amp_c(u,s,t)\;.
\end{equation}
However, this relation is only valid at leading order in isospin breaking and can not be used at $\ord(e^2\delta)$. This is most easily seen by the fact that e.g.\ photon loop contributions do not respect it. Furthermore, an explicit check using the LO amplitudes~\eqref{eqn:cloamp} and~\eqref{eqn:nloamp} shows that this relation is only valid except for terms proportional to $\delta\times(3s_0^c-3s_0^n)=\delta\times2\Delta\mpi$. Thus both decay channels have to be calculated separately.

At next-to-leading chiral order the $\etp$ decay amplitudes receive various strong and electromagnetic contributions. Besides renormalization effects and $\eta\pi^0$ mixing at NLO, both strong and electromagnetic NLO low-energy constants enter the calculation via the tree diagram with a vertex from the NLO Lagrangian. The $I=0$ rescattering of intermediate pions is particularly important since it gives rise to about half of the total NLO corrections, cf.\ Ref.~\cite{gleta}. For the charged decay, the infrared~(IR) divergences due to virtual-photon loops are canceled by including real-photon radiation (bremsstrahlung) in the soft-photon approximation at the amplitude-squared level. The exchange of a virtual photon in the final state between the charged pions leads to a triangle loop function which has some interesting features: both the real and the imaginary part are IR-divergent; while the IR divergence in the real part is cancelled against bremsstrahlung contributions, the imaginary part can be resummed in the Coulomb phase; furthermore it contains a kinematical singularity at threshold $s=4\mpc$, the Coulomb pole. The kinematical singularities in the radiative corrections, i.e.\ the Coulomb pole and the kinematical bremsstrahlung singularity at $s=(M_\eta-M_{\pi^0})^2$, are part of the universal soft-photon corrections, see e.g.\ Ref.~\cite{isidorisoftphoton}, that are usually already applied in the analysis of the experimental data in order to perform a meaningful fit of the Dalitz plot distribution. Hence we omit the unobservable Coulomb phase and subtract the Coulomb pole and the kinematical Bremsstrahlung singularity from the squared amplitude for the derivation of the numerical results.

We have checked both the charged and the neutral amplitude in several ways: they are finite and renormalization-scale independent and, by use of relation~\eqref{eqn:isoamprel}, both amplitudes reduce to the results of GL~\cite{gleta} at $\ord(\delta)$ and of BKW~\cite{bkw} at $\ord(e^2)$. In addition, we have checked explicitly that at the soft-pion point~\cite{sutherland} the new electromagnetic corrections are relatively suppressed in comparison with the BKW corrections only by $\delta/\hat{m}\approx2/3$, as expected.

%================================================================================
\section{Results}
%\label{sec:results}
%================================================================================

It is well known that one has to go beyond one-loop chiral order to obtain a phenomenologically successful representation of the $\etp$ decay amplitudes. Since we focus on the electromagnetic contributions anyway, all results are given as relative corrections to the results obtained via the purely strong amplitude at $\ord(\delta)$ which corresponds precisely to the GL amplitude. Furthermore we only consider uncertainties in the electromagnetic corrections and diregard higher-order hadronic corrections. Hence the purely electromagnetic corrections at $\ord(e^2)$ corresponding to the BKW amplitude as well as the new mixed corrections at $\ord(e^2\delta)$ are successively added to the strong amplitude and their uncertainties are obtained by variation of the rather unknown electromagnetic low-energy constants.

\begin{figure}
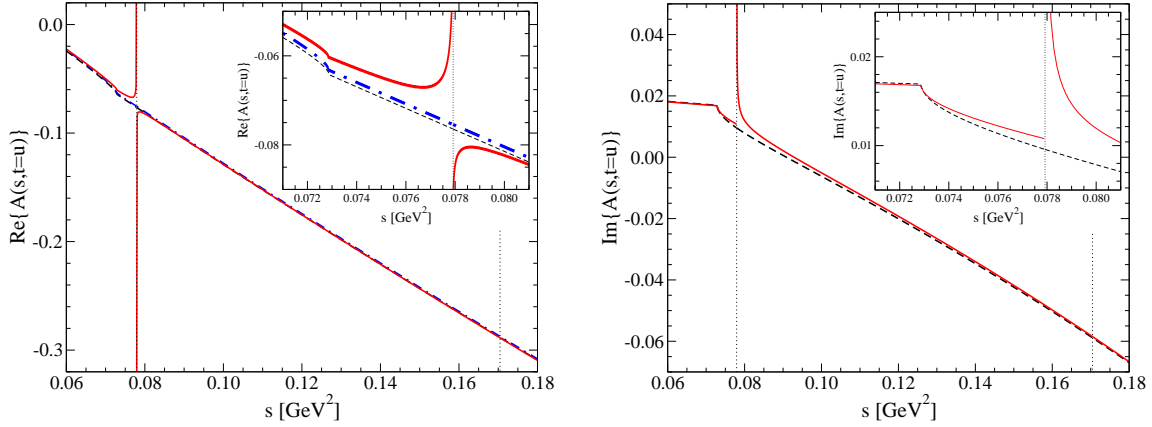

 \centering
 \includegraphics[width=0.48\linewidth]{campre.eps} \hfill \includegraphics[width=0.48\linewidth]{campim.eps}
 \caption{Real and imaginary parts of the charged amplitudes GL (dashed/black), BKW (dot-dashed/blue), and DKM (full/red) for $t=u$. The inserts show the region close to the two-pion thresholds. The line widths in the real part indicate the error bands.}
 \label{fig:campzo}
\end{figure}

\begin{figure}
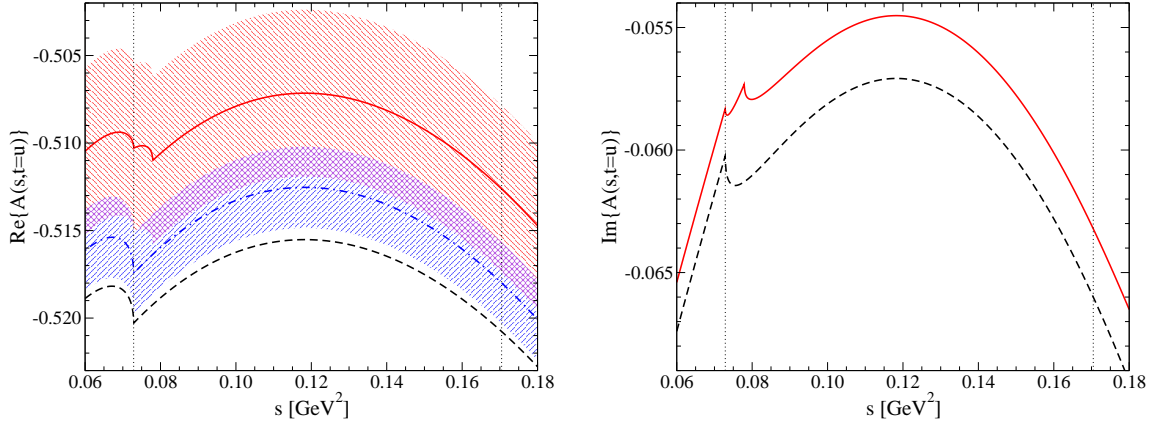

 \centering
 \includegraphics[width=0.48\linewidth]{nampre.eps} \hfill \includegraphics[width=0.48\linewidth]{nampim.eps}
 \caption{Real and imaginary parts of the neutral amplitudes GL (dashed/black), BKW (dot-dashed/blue), and DKM (full/red) for $t=u$. The hatched regions denote the error bands.}
 \label{fig:namp}
\end{figure}

A comparison of the different electromagnetic contributions at the amplitude level is given in Figs.~\ref{fig:campzo} and~\ref{fig:namp}, where the real and the imaginary parts of the amplitudes at different orders in isospin breaking (GL, BKW and DKM) are shown separately for each decay channel. The amplitudes are plotted along the lines with $t=u$, the vertical dotted lines show the limits of the corresponding physical regions. The BKW corrections are purely real for both decay channels, as no pion rescattering diagrams contribute at that particular order, and thus the imaginary parts of GL and BKW coincide with each other. Furthermore, all imaginary parts are independent of any low-energy constants and therefore plotted without an error range.
The \textit{charged} decay amplitude exhibits the kinematical singularities at threshold $s=4\mpc$ (Coulomb pole \& phase) that are retained here for illustrative reasons, while the IR divergences in the amplitude are cured by hand. By our choice of the neutral pion mass for the isospin limit, the threshold cusp in the GL amplitude is artificially removed from the physical threshold energy to $s=4\mpn$.
Since the LO \textit{neutral} decay amplitude is constant and also at NLO the dependence on $s$ is weak, the overall variation and thereby the scale of Fig.~\ref{fig:namp} is very small. Thus the electromagnetic uncertainties as well as the expected cusp at the energy of the $\pi^+\pi^-$ threshold inside the physical region at $\ord(e^2\delta)$ are clearly visible. Furthermore the figure shows that the size of the DKM contributions is comparable to or even larger than the BKW contributions.

\begin{figure}
 \centering
 \includegraphics[scale=0.9]{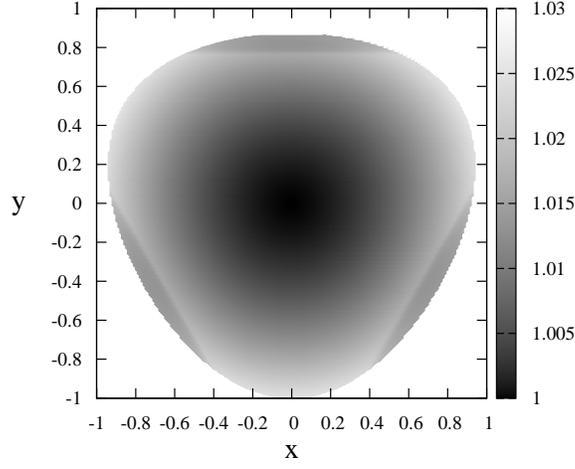}
 \caption{$\etpn$ Dalitz plot distribution corresponding to the DKM amplitude.}
 \label{fig:3Dcusp}
\end{figure}

The Dalitz plot for the charged decay is described in terms of the symmetrized coordinates
\begin{equation}
 x=\sqrt{3}\frac{T_+-T_-}{Q_c}=\frac{\sqrt{3}(u-t)}{2M_\eta Q_c}\;, \qquad y=\frac{3T_0}{Q_c}-1=\frac{3\bigl[(M_\eta-M_{\pi^0})^2-s\bigr]}{2M_\eta Q_c}-1\;,
\end{equation}
whereas for the neutral decay it is convenient to use the fully symmetrized coordinate
\begin{equation}
 z=\frac{2}{3}\sum_{i=1}^{3}\biggl(\frac{3T_i}{Q_n}-1\biggr)^2=x^2+y^2\;.
\end{equation}
Here $Q_c=T_0+T_++T_-=M_\eta-2M_{\pi^\pm}-M_{\pi^0}$ and $Q_n=T_1+T_2+T_3=M_\eta-3M_{\pi^0}$, where the $T_i$ are the kinetic energies of the respective pions in the $\eta$ rest frame. In order to obtain the Dalitz plot parameters the squared absolute values of the decay amplitudes are expanded around the center of the corresponding Dalitz distribution according to the standard parameterizations
\begin{align}
 \begin{split}
  |\amp_c(x,y)|^2 &= |\mathcal{N}_c|^2\bigl\{1+ay+by^2+dx^2+...\bigr\}\;,
  \\
  |\amp_n(x,y)|^2 &= |\mathcal{N}_n|^2\bigl\{1+2\alpha z+...\bigr\}\;.
 \end{split}
\end{align}
For the \textit{charged} channel, the purely strong results and the successive electromagnetic relative corrections are given in Tab.~\ref{tab:cdalitzpar}. The normalization gets reduced and the slopes tend to increase. All corrections are at the percent level, but the BKW corrections do not represent a valid estimate of the dominant electromagnetic corrections.
\begin{table}
 \centering
 \newcommand{\tbk}{\hspace{-4mm}}
 \newcommand{\tpm}{\tbk$\pm$}
 \renewcommand{\arraystretch}{1.2}
 \begin{tabular}{|crclrclrclrcl|}
  \hline
  &\multicolumn{3}{c}{$|\mathcal{N}_{c}|^2$}&\multicolumn{3}{c}{$a$}&\multicolumn{3}{c}{$b$}&\multicolumn{3}{c|}{$d$}\\
  \hline
  GL          & \multicolumn{3}{c}{$0.0325$} & \multicolumn{3}{c}{$-1.279$} & \multicolumn{3}{c}{$0.396$} & \multicolumn{3}{c|}{$0.0744$} \\
  $\Delta$BKW & $(-1.1 $&\tpm&\tbk$0.9)\%$   & $(+0.6$&\tpm&\tbk$0.1)\%$    & $(+1.4$&\tpm&\tbk$0.2)\%$   & $(+1.5$ &\tpm&\tbk$0.5)\%$ \\
  $\Delta$DKM & $(-2.4 $&\tpm&\tbk$0.7)\%$   & $(+0.7$&\tpm&\tbk$0.4)\%$    & $(+1.5$&\tpm&\tbk$0.7)\%$   & $(+4.4$ &\tpm&\tbk$0.4)\%$ \\
  \hline
 \end{tabular}
 \renewcommand{\arraystretch}{1.0}
 \caption{Dalitz normalization and slopes for $\etpc$ resulting from the GL amplitude and relative electromagnetic corrections due to BKW and DKM amplitudes.}
 \label{tab:cdalitzpar}
\end{table}
The Dalitz plot parameters for the \textit{neutral} channel are shown in Tab.~\ref{tab:ndalitzpar}. By DKM(w/o cusp) we denote a fit to the inner part of the Dalitz plot with $z\leq z_0$ such that the border region from the cusp outward is excluded, since the cusp is incompatible with a simple polynomial fit. As for the charged decay channel, the normalization gets reduced by a few percent. Here, the corrections of $\ord(e^2\delta)$ are even bigger than those of $\ord(e^2)$. Note that the cusp effect leads to the single biggest modification of any Dalitz plot parameter: trying to fit the cusp with the polynomial function reduces $\alpha$ by 4\% (compare $\Delta$DKM to $\Delta$BKW), while excluding the cusp region increases it again by more than 5\%. The significance of this non-analytic structure is also reflected in the fit quality as quantified by the quoted $\chi^2/\ndf$ values.
\begin{table}
 \centering
 \newcommand{\tbk}{\hspace{-4mm}}
 \newcommand{\tpm}{\tbk$\pm$}
 \renewcommand{\arraystretch}{1.2}
 \begin{tabular}{|crclrclc|}
  \hline
  &\multicolumn{3}{c}{$|\mathcal{N}_{n}|^2$}&\multicolumn{3}{c}{$10^2\times\alpha$}&$\chi^2/\ndf$\\
  \hline
  GL                    & \multicolumn{3}{c}{$ 0.269$} & \multicolumn{3}{c}{$ 1.27$} & $\equiv1$ \\
  $\Delta$BKW           & $(-1.1$&\tpm&\tbk$0.9)\%$    & $(+3.7$&\tpm&\tbk$0.5)\%$   & $0.99$    \\
  $\Delta$DKM           & $(-3.3$&\tpm&\tbk$1.8)\%$    & $(-0.2$&\tpm&\tbk$1.0)\%$   & $6.20$    \\
  $\Delta$DKM(w/o cusp) & $(-3.3$&\tpm&\tbk$1.8)\%$    & $(+5.0$&\tpm&\tbk$1.1)\%$   & $0.35$    \\
  \hline
 \end{tabular}
 \renewcommand{\arraystretch}{1.0}
 \caption{Dalitz normalization and slopes for $\etpn$ resulting from the GL amplitude and relative electromagnetic corrections due to BKW and DKM amplitudes. For details, see main text.}
 \label{tab:ndalitzpar}
\end{table}
This behavior can easily be understood qualitatively by looking at the Dalitz plot distribution at $\ord(e^2\delta)$ presented in Fig.~\ref{fig:3Dcusp}. The distribution can be thought of as a bowl with three symmetrical handles where the surface bends down due to the cusp at constant $\pi^+\pi^-$ threshold in $s$, $t$ and $u$ each.

The decay widths that can easily be calculated from the Dalitz plot distributions are given in Tab.~\ref{tab:dwbrqmr}. For comparison we also show the result for the charged amplitude without subtraction of the universal soft-photon corrections, named DKM(uc). For the neutral decay, the shifts of the width are completely dominated by the corrections of the Dalitz normalizations and the DKM corrections are twice as large as the BKW corrections. When looking at the neutral-to-charged branching ratio, the difference between the BKW and the DKM result becomes even larger, since at $\ord(e^2)$ the relative shifts nearly cancel in the ratio. Finally, by use of the (approximate) relation $\Gamma_{c/n}\propto Q^{-4}$ one can read off the corrections to the quark mass ratio $Q$ which are also quoted in Tab.~\ref{tab:dwbrqmr}. This relation does not hold for the BKW contributions, but the additional error in factorizing $Q^{-2}$ in the complete amplitude can be safely neglected. In order to purify a real measurement (which includes electromagnetic effects) such that the purely strong value for $Q$ can be extracted, the opposite shift has to be applied. Note that there are also ongoing theoretical efforts to obtain $Q$ from a dispersive treatment of $\etp$ decays~\cite{lanz}.
\begin{table}
  \centering
 \newcommand{\tbk}{\hspace{-4mm}}
 \newcommand{\tpm}{\tbk$\pm$}
 \renewcommand{\arraystretch}{1.2}
 \begin{tabular}{|crclrcl|crcl|}
  \hline
                              & \multicolumn{3}{c}{$\etpc$}    & \multicolumn{3}{c|}{$\etpn$}    &  \multicolumn{4}{c|}{$r=\Gamma_n/\Gamma_c$}     \\
  \hline
  $\Gamma^\tn{GL}$            & \multicolumn{3}{c}{$154.5\ev$} & \multicolumn{3}{c|}{$222.8\ev$} & $r^\tn{GL}$             & \multicolumn{3}{c|}{$1.442$} \\
  $\Delta\Gamma^\tn{BKW}$     & $(-1.0$&\tpm&\tbk$0.9)\%$      & $(-1.1$&\tpm&\tbk$0.9)\%$       & $\Delta r^\tn{BKW}$     & $(-0.1$&\tpm&\tbk$1.2)\%$    \\
  $\Delta\Gamma^\tn{DKM}$     & $(-1.9$&\tpm&\tbk$0.5)\%$      & $(-3.3$&\tpm&\tbk$1.8)\%$       & $\Delta r^\tn{DKM}$     & $(-1.4$&\tpm&\tbk$1.8)\%$    \\
  $\Delta\Gamma^\tn{DKM(uc)}$ & $(-1.0$&\tpm&\tbk$0.5)\%$      &        &    &                   & $\Delta r^\tn{DKM(uc)}$ & $(-2.3$&\tpm&\tbk$1.8)\%$    \\
  \hline
  $\Delta Q^\tn{BKW}$         & $(+0.24$&\tpm&\tbk$0.22)\%$    & $(+0.28$&\tpm&\tbk$0.22)\%$ & & & & \\
  $\Delta Q^\tn{DKM}$         & $(+0.48$&\tpm&\tbk$0.12)\%$    & $(+0.84$&\tpm&\tbk$0.46)\%$ & \multicolumn{4}{c|}{$\Gamma_{c/n}\propto Q^{-4}$} \\
  $\Delta Q^\tn{DKM(uc)}$     & $(+0.24$&\tpm&\tbk$0.12)\%$    &         &    &              & & & & \\
  \hline
 \end{tabular}
 \renewcommand{\arraystretch}{1.0}
 \caption{Decay widths, branching ratio and quark mass ratio $Q$.}
 \label{tab:dwbrqmr}
\end{table}

%================================================================================
\section{Summary \& Outlook}
%\label{sec:summary}
%================================================================================

Although electromagnetic contributions to $\etp$ decays are small in general, they ought to be accounted for in high-precision studies. The effects at isospin breaking $\ord(e^2\delta)$ are as large as the effects at $\ord(e^2)$.

Since the present calculation is performed in ChPT at NLO and hence includes the rescattering effect leading to the cusp only at LO in the quark mass expansion, it is not suited to serve for a precision extraction of $\pi\pi$ scattering lengths from the cusp effect, but rather illustrates the phenomenon. The theoretical framework perfectly suited for such an extraction is non-relativistic effective field theory~\cite{colangelocusp,bisseggercusp,bisseggerradcorr,gullstrom,schneider}. The present calculation is in a sense dual to the non-relativistic calculation as it aims at predicting electromagnetic effects in those parts of the amplitude that are merely parameterized in the latter.%\\[-2mm]

\acknowledgments
I am very grateful to my collaborators and thank the organizers for making this exciting conference possible.
Partial financial support by the Helmholtz Association through funds provided to the virtual institute ``Spin and strong QCD'' (VH-VI-231), by the EU Integrated Infrastructure Initiative Hadron Physics Project (contract number RII3--CT--2004--506078), and by DFG (SFB/TR 16, ``Subnuclear Structure of Matter'') is gratefully acknowledged.

%\appendix

\end{document}